\documentclass[aps,twocolumn,prl,superscriptaddress,showpacs]{revtex4}

\usepackage{graphicx, amsmath}

\renewcommand{\bar}[1]{\overline{#1}{}}

\newcommand{\de}{\Delta E}
\newcommand{\dm}{\Delta m}
\newcommand{\dt}{\Delta t}
\newcommand{\mbc}{M_{\rm bc}}
\newcommand{\bb}{B{\bar B}}
\newcommand{\qq}{q{\bar q}}
\newcommand{\ks}{K^0_S}
\newcommand{\ds}{D^{*}}
\newcommand{\bdnpn}{\bar{B}^0\to D h^0}
\newcommand{\bdspn}{\bar{B}^0\to D^* h^0}
\newcommand{\kspipi}{\ks\pi^+\pi^-}
\newcommand{\dnkspipi}{D\to\ks\pi^+\pi^-}
\newcommand{\bdbkspipi}{\bar{B}^0\to D[ \kspipi] h^0}

\newcommand{\sinphi}{\sin 2\phi_1  = 0.725 \pm 0.037}
\newcommand{\sinres}{\sin 2\phi_1=0.78\pm 0.44\pm 0.22}
\newcommand{\cosres}{\cos 2\phi_1=1.87^{+0.40+0.22}_{-0.53-0.32}}
\newcommand{\clres}{98.3\%}
\begin{document}

\title{\quad\\[0.5cm] \boldmath
Measurement of the CKM parameter $\cos 2\phi_1$ using time-dependent
Dalitz analysis of $\bdbkspipi$
}
\affiliation{Budker Institute of Nuclear Physics, Novosibirsk}
\affiliation{Chiba University, Chiba}
\affiliation{University of Cincinnati, Cincinnati, Ohio 45221}
\affiliation{University of Hawaii, Honolulu, Hawaii 96822}
\affiliation{High Energy Accelerator Research Organization (KEK), Tsukuba}
\affiliation{Institute of High Energy Physics, Vienna}
\affiliation{Institute of High Energy Physics, Protvino}
\affiliation{Institute for Theoretical and Experimental Physics, Moscow}
\affiliation{J. Stefan Institute, Ljubljana}
\affiliation{Kanagawa University, Yokohama}
\affiliation{Korea University, Seoul}
\affiliation{Kyungpook National University, Taegu}
\affiliation{Swiss Federal Institute of Technology of Lausanne, EPFL, Lausanne}
\affiliation{University of Maribor, Maribor}
\affiliation{University of Melbourne, Victoria}
\affiliation{Nagoya University, Nagoya}
\affiliation{Nara Women's University, Nara}
\affiliation{National Central University, Chung-li}
\affiliation{National United University, Miao Li}
\affiliation{Department of Physics, National Taiwan University, Taipei}
\affiliation{H. Niewodniczanski Institute of Nuclear Physics, Krakow}
\affiliation{Nippon Dental University, Niigata}
\affiliation{Niigata University, Niigata}
\affiliation{Osaka City University, Osaka}
\affiliation{Osaka University, Osaka}
\affiliation{Panjab University, Chandigarh}
\affiliation{Peking University, Beijing}
\affiliation{Princeton University, Princeton, New Jersey 08544}
\affiliation{University of Science and Technology of China, Hefei}
\affiliation{Seoul National University, Seoul}
\affiliation{Shinshu University, Nagano}
\affiliation{Sungkyunkwan University, Suwon}
\affiliation{University of Sydney, Sydney NSW}
\affiliation{Tata Institute of Fundamental Research, Bombay}
\affiliation{Toho University, Funabashi}
\affiliation{Tohoku Gakuin University, Tagajo}
\affiliation{Tohoku University, Sendai}
\affiliation{Department of Physics, University of Tokyo, Tokyo}
\affiliation{Tokyo Institute of Technology, Tokyo}
\affiliation{Tokyo Metropolitan University, Tokyo}
\affiliation{University of Tsukuba, Tsukuba}
\affiliation{Virginia Polytechnic Institute and State University, Blacksburg, Virginia 24061}
\affiliation{Yonsei University, Seoul}
 \author{P.~Krokovny}\affiliation{High Energy Accelerator Research Organization (KEK), Tsukuba} 
   \author{K.~Abe}\affiliation{High Energy Accelerator Research Organization (KEK), Tsukuba} 
   \author{K.~Abe}\affiliation{Tohoku Gakuin University, Tagajo} 
   \author{I.~Adachi}\affiliation{High Energy Accelerator Research Organization (KEK), Tsukuba} 
   \author{H.~Aihara}\affiliation{Department of Physics, University of Tokyo, Tokyo} 
   \author{D.~Anipko}\affiliation{Budker Institute of Nuclear Physics, Novosibirsk} 
   \author{K.~Arinstein}\affiliation{Budker Institute of Nuclear Physics, Novosibirsk} 
   \author{Y.~Asano}\affiliation{University of Tsukuba, Tsukuba} 
   \author{V.~Aulchenko}\affiliation{Budker Institute of Nuclear Physics, Novosibirsk} 
   \author{T.~Aushev}\affiliation{Institute for Theoretical and Experimental Physics, Moscow} 
% ...
\collaboration{The Belle Collaboration}

\begin{abstract} 
  \noindent
We present a measurement of the angle $\phi_1$ of 
the CKM Unitarity Triangle using time-dependent Dalitz analysis of
$D\to\kspipi$ decays produced in neutral $B$ meson decay to a neutral
$D$ meson and a light meson ($\bar{B}^0\to D^{(*)} h^0$).
The method allows a direct extraction of $2\phi_1$ and, therefore,
helps to resolve the ambiguity between $2\phi_1$ and $\pi-2\phi_1$
in the measurement of $\sin 2\phi_1$. We obtain 
$\sinres$ and $\cosres$. The sign of $\cos 2\phi_1$ is determined to be
positive at $\clres$ C.L.
\end{abstract}


\maketitle


Precise determination of the Cabibbo-Kobayashi-Maskawa (CKM) 
matrix elements~\cite{ckm} is important to 
check the consistency of the Standard Model (SM) and search for new physics.
The value of $\sin 2\phi_1$, where $\phi_1$ is one of the angles of 
the Unitarity Triangle, is now measured with high precision: 
$\sinphi$~\cite{PDG,sin2phi1}. 
This leads to four solutions in $\phi_1$: $23^\circ$, $67^\circ$,
$(23+180)^\circ$, and $(67+180)^\circ$. Resolution of this
ambiguity has been attempted using time-dependent angular analysis
in the ${B}^0 \to J/\psi K^{*0} (K_S^0 \pi^0)$ decay. This technique 
provides a measurement of $\cos 2\phi_1$ and therefore helps to distinguish 
between the solutions at $23^\circ$ and 
$67^\circ$~\cite{kstar_babar, kstar_belle}.

A new technique based on the analysis of $\bdbkspipi$ has been 
recently suggested~\cite{bgk}.
Here we use $h^0$ to denote light neutral mesons, 
$\pi^0$, $\eta$ and $\omega$.
The neutral $D$ meson is reconstructed in the $\kspipi$ decay mode,
its resonant substructure has been measured~\cite{dkpp_cleo, anton}.


Consider a neutral $B$ meson
that is known to be a $\bar{B}^0$ at time $t_{\rm tag}$.
At another time, $t_{\rm sig}$, its state is given by
\begin{eqnarray}
  \label{eq:b0bar_evo}
&&\left| \bar{B}^0(\dt) \right> = 
  e^{-\left|\dt\right|/2\tau_{B^0}}\times\\\nonumber
&&\Bigl(
    \left| \bar{B}^0 \right> \cos(\dm\dt/2) - 
    i \frac{p}{q} \left| B^0 \right> \sin(\dm\dt/2)
  \Bigr),
\end{eqnarray}
where $\dt = t_{\rm sig} - t_{\rm tag}$, 
$\tau_{B^0}$ is the average lifetime of the $B^0$ meson,
$\dm$, $p$ and $q$ are parameters of $B^0$-$\bar{B}^0$ mixing.
Here we have assumed $CPT$ invariance and 
neglected terms related to the lifetime difference of neutral $B$ mesons.
In the SM, $| q/p | = 1$ to a good approximation,
and, in the usual phase convention, 
${\rm arg}(p/q) = 2 \phi_1$.

\begin{figure}
  \includegraphics[width=0.35\textwidth]{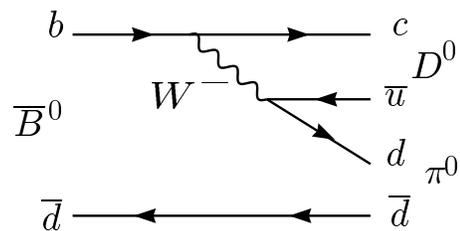}
  \caption{
    \label{diag_fav}
    Diagram for the dominant color-suppressed amplitude 
    for $\bar{B}^0 \to D\pi^0$.
  }
\end{figure}

The $B\to D h^0$ decay amplitude is dominated by the CKM favored 
$b\to c\bar{u}d$ diagram as shown in Fig.~\ref{diag_fav}, 
with roughly a 2\% contribution
from the CKM suppressed $b\to u\bar{c}d$ diagram. Ignoring the latter, 
a neutral $D$ meson produced in a $\bar{B}^0$ decay is a $D^0$, 
while that produced in a $B^0$ decay is a $\bar{D}^0$.
The $D$ meson state produced at time $\dt$ is then given by
$  \left| D^0 \right> \cos(\dm\dt/2) - 
  i e^{2i\phi_1}\xi_{h^0} (-1)^l \left| \bar{D}^0 \right> \sin(\dm\dt/2)$,
where we use $\xi_{h^0}$ to denote the $CP$ eigenvalue of $h^0$,
and $l$ gives the orbital angular momentum in the $Dh^0$ 
system. In the case of $\bar{B}^0 \to D^*h^0$, 
an additional factor arises due to the $CP$ properties
of the particle emitted in the $D^*$ decay 
(either $D^* \to D\pi^0$ or $D^* \to D\gamma$)~\cite{dstar}.

We follow Ref.~\cite{anton} and describe the amplitude for a 
$\bar{D}^0\to\kspipi$ decay as $f(m_+^2,m_-^2)$, 
where $m_+^2$ and $m_-^2$ are the squares of the
two-body invariant masses of the $\ks\pi^+$ and $\ks\pi^-$ combinations.
Assuming no $CP$ violation in the neutral $D$ meson system,
the amplitude for a $D^0$ decay is then given by $f(m_-^2,m_+^2)$.
The time-dependent Dalitz plot density is defined by
\begin{eqnarray}
  \label{eq:pdf}\nonumber
&&P(m_+^2, m_-^2, \dt, q_B) =
\frac{e^{-|\dt|/\tau_{B^0}}}{8\tau_{B^0}}
\frac{F(m_+^2,m_-^2)}{2N}\Bigl( 1 + q_B\times\\\nonumber
&&\bigl\{{\mathcal A} (m_-^2, m_+^2)\cos(\dm\dt) 
+ {\mathcal S}(m_-^2, m_+^2)\sin(\dm\dt)\bigr\}\Bigr),\\\nonumber
&&{\mathcal A} = 
(|f(m_-^2,m_+^2)|^2 - |f(m_+^2,m_-^2)|^2)/F(m_+^2,m_-^2),\\\nonumber
&&{\mathcal S} = \frac{-2\xi_{h^0}(-1)^l 
Im\{f(m_-^2,m_+^2)f^*(m_+^2,m_-^2)
e^{2i\phi_1}\}}{F(m_+^2,m_-^2)},\\\nonumber
&&F = |f(m_-^2,m_+^2)|^2 + |f(m_+^2,m_-^2)|^2,\\
&&N = \int{|f(m_-^2,m_+^2)|^2 dm^2_+ dm^2_-},
\end{eqnarray}
where the $b$-flavor charge is $q_B=+1$ ($-1$) when the tagging $B$ meson
is a $B^0$ ($\bar{B}^0$).
Thus the phase $2\phi_1$ can be extracted from 
a time-dependent Dalitz plot fit to $B^0$ and $\bar{B}^0$ data
if $f(m_+^2,m_-^2)$ is known.
Note that this formulation assumes that there is no direct $CP$
violation in the $B$ decay amplitudes.


This analysis is based on $386\times 10^6$ $B\bar{B}$ events collected 
with the
Belle detector at the asymmetric energy $e^+e^-$ collider~\cite{KEKB}.
The Belle detector has been described elsewhere~\cite{belle}.
We reconstruct the decays $\bdnpn$ for $h^0 = \pi^0, \eta$ and 
$\omega$ and $\bdspn$ for $h^0 = \pi^0$ and $\eta$.

Charged tracks are selected based on the
number of hits and impact parameter relative to the
interaction point (IP).
To reduce combinatorial background, a transverse momentum of at least 
0.1~GeV$/c$ is required of each track. 
All charged tracks that are not positively identified as electrons 
are treated as pions.

Neutral kaons are reconstructed via the decay $\ks\to\pi^+\pi^-$.
The $\pi\pi$ invariant mass is required to be within 9~MeV$/c^2$
($\sim 3\sigma$) of the $K^0$ mass, and the displacement of the 
$\pi^+\pi^-$ vertex from the IP in the transverse ($r$-$\varphi$) 
plane is required to have a magnitude between 0.2~cm and 20~cm and
a direction that agrees within 0.2 
radians with the combined momentum of the two pions.

Photon candidates are selected from calorimeter showers not associated
with charged tracks.
An energy deposition of at least 50~MeV and a photon-like shape are 
required for each candidate.
A pair of photons with an invariant mass 
within 12~MeV$/c^2$ ($2.5\sigma$) of the $\pi^0$ mass is 
considered as a $\pi^0$ candidate.

We reconstruct neutral $D$ mesons in the $\kspipi$ decay channel
and require the invariant mass to be within 15~MeV$/c^2$ ($2.5\sigma$)
of the nominal $D^0$ mass.
$D^{*0}$ candidates are reconstructed in the $D^0\pi^0$ decay channel.
The mass difference between $D^{*0}$ and $D^0$ candidates is required 
to be within 3~MeV$/c^2$ of the expected value ($3\sigma$).
$\omega$ candidates are reconstructed in the $\pi^+\pi^-\pi^0$ decay 
channel. Their invariant mass is required to be within 20~MeV$/c^2$ 
($2.5~\Gamma$) of the $\omega$ mass. 
We define the angle $\theta_\omega$ between the normal to the 
$\omega$ decay plane and opposite of the $B$ direction in the rest
frame of $\omega$
and require $|\cos \theta_\omega|>0.3$.
We reconstruct  $\eta$ candidates in the $\gamma\gamma$ and 
$\pi^+\pi^-\pi^0$ final states and require the invariant mass to be 
within 10 and 30~MeV$/c^2$ ($2.5\sigma$) of the $\eta$ mass, 
respectively. The photon energy threshold for the prompt $\pi^0$ 
and $\eta$ candidates coming from $B$ decays is
increased to 200~MeV in order to reduce combinatorial background.
We remove $\eta$ candidates if either of the daughter photons can be 
combined with any other photon with $E_\gamma>100$~MeV to form a 
$\pi^0$ candidate.

We combine either $D$ and $h^0=\{\pi^0,\omega,\eta\}$ 
or $\ds$ and $h^0=\{\pi^0,\eta\}$ to form $B$ mesons. 
Signal candidates are identified by their 
energy difference in the center-of-mass system of the 
$\Upsilon$(4S) (CM), 
\mbox{$\de=(\sum_iE_i)-E_{\rm beam}$}, and the beam-energy constrained mass, 
$\mbc=\sqrt{E_{\rm beam}^2-(\sum_i\vec{p}_i)^2}$, where $E_{\rm beam}$ 
is the beam energy and $\vec{p}_i$ and $E_i$ are the momenta and 
energies of the decay products of the $B$ meson in the CM frame. 
The masses of $\pi^0$, $\eta$ and $D^{(*)}$ candidates are constrained
to their nominal values to improve $\de$ resolution.
We select events with $\mbc>5.2$~GeV$/c^2$ and $|\de|<0.3$~GeV,
and define the signal region to be
$5.272$~GeV$/c^2<\mbc<5.287$~GeV$/c^2$, 
$-0.1~{\rm GeV}<\de<0.06$~GeV ($\pi^0$, $\eta\to\gamma\gamma$) 
or $|\de|<0.03$~GeV ($\omega$, $\eta\to\pi^+\pi^-\pi^0$).
In cases with more than one candidate in an event, the one with
$D$ and $h^0$ masses closest to the nominal values
is chosen.

To suppress the large combinatorial background dominated by 
the two-jet-like $e^+e^-\to\qq$ continuum 
process, variables that characterize the event topology are used. 
We require $|\cos\theta_{\rm thr}|<0.80$, where $\theta_{\rm thr}$ is 
the angle between the thrust axis of the $B$ candidate and that of the 
rest of the event. This requirement eliminates 77\% of the continuum 
background and retains 78\% of the signal. We also construct a 
Fisher discriminant, ${\cal F}$, which is based on the production
angle of the $B$ candidate, the angle of the $B$ candidate thrust axis 
with respect to the beam axis, and nine parameters that characterize 
the momentum flow in the event relative to the $B$ candidate thrust 
axis in the CM frame~\cite{fisher}. We impose a requirement on 
${\cal{F}}$ that rejects 67\% of the remaining continuum background 
and retains 83\% of the signal. 

\begin{figure*}
\includegraphics[width=0.24\textwidth] {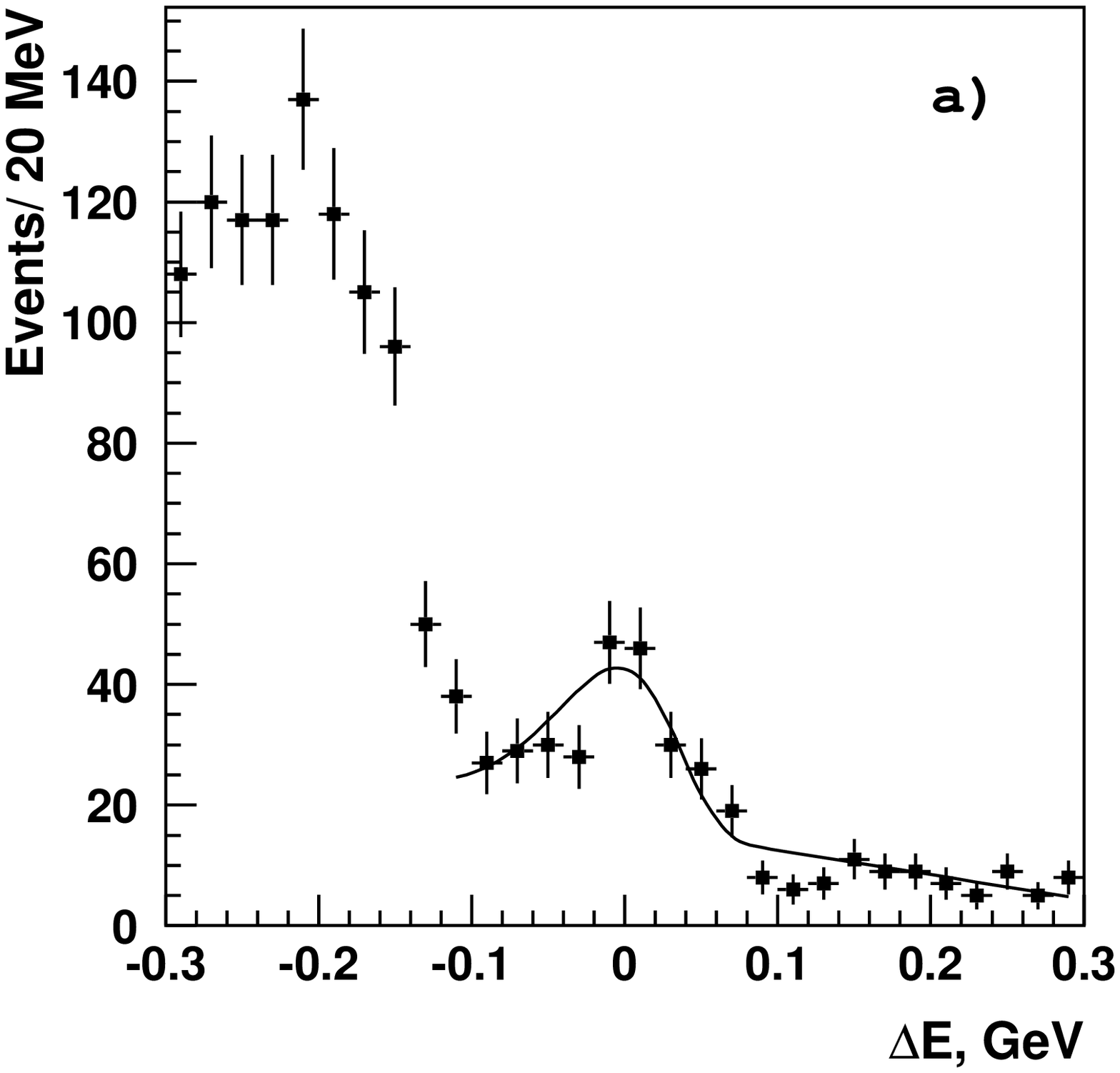}
\includegraphics[width=0.24\textwidth] {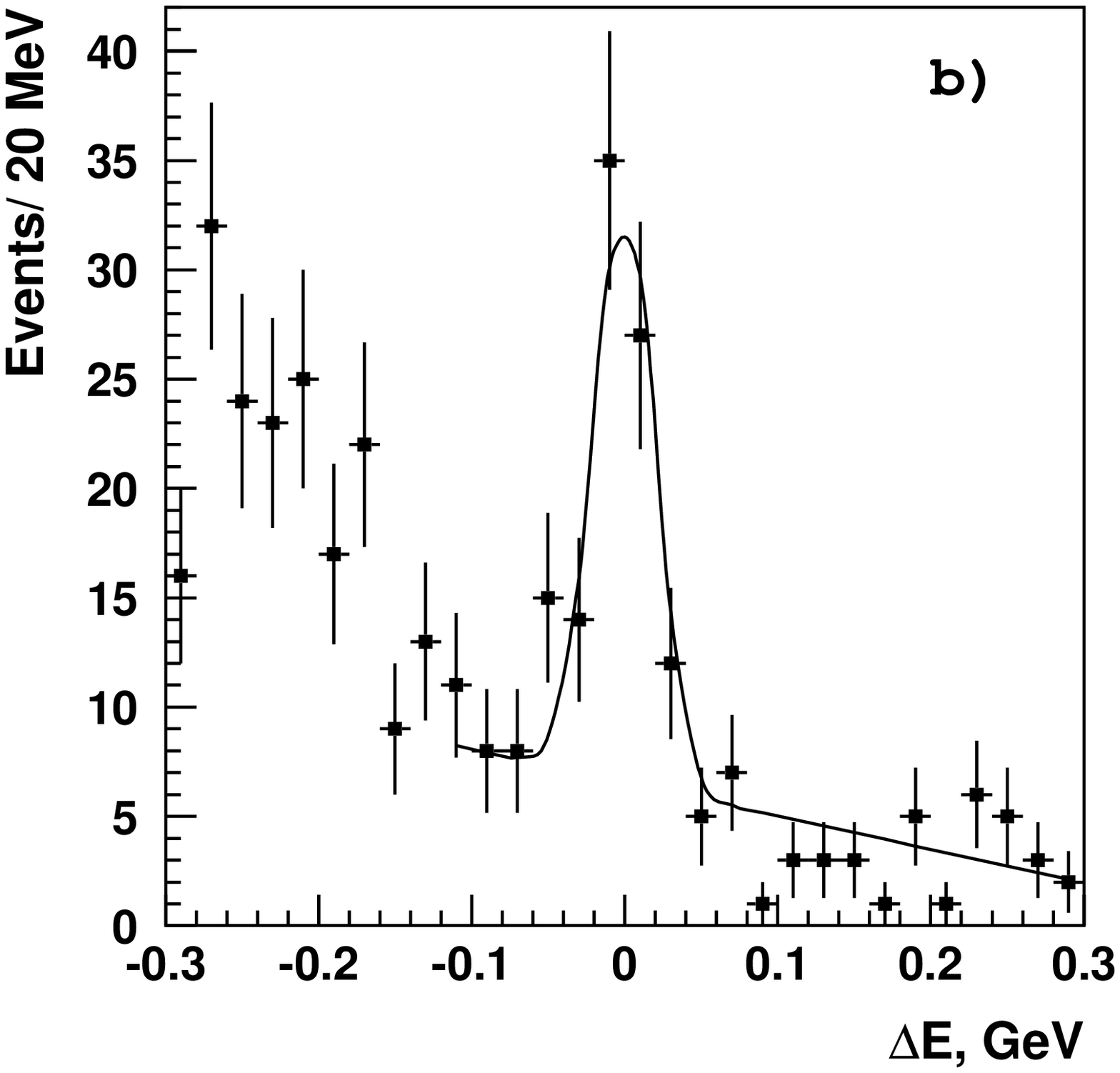}
\includegraphics[width=0.24\textwidth] {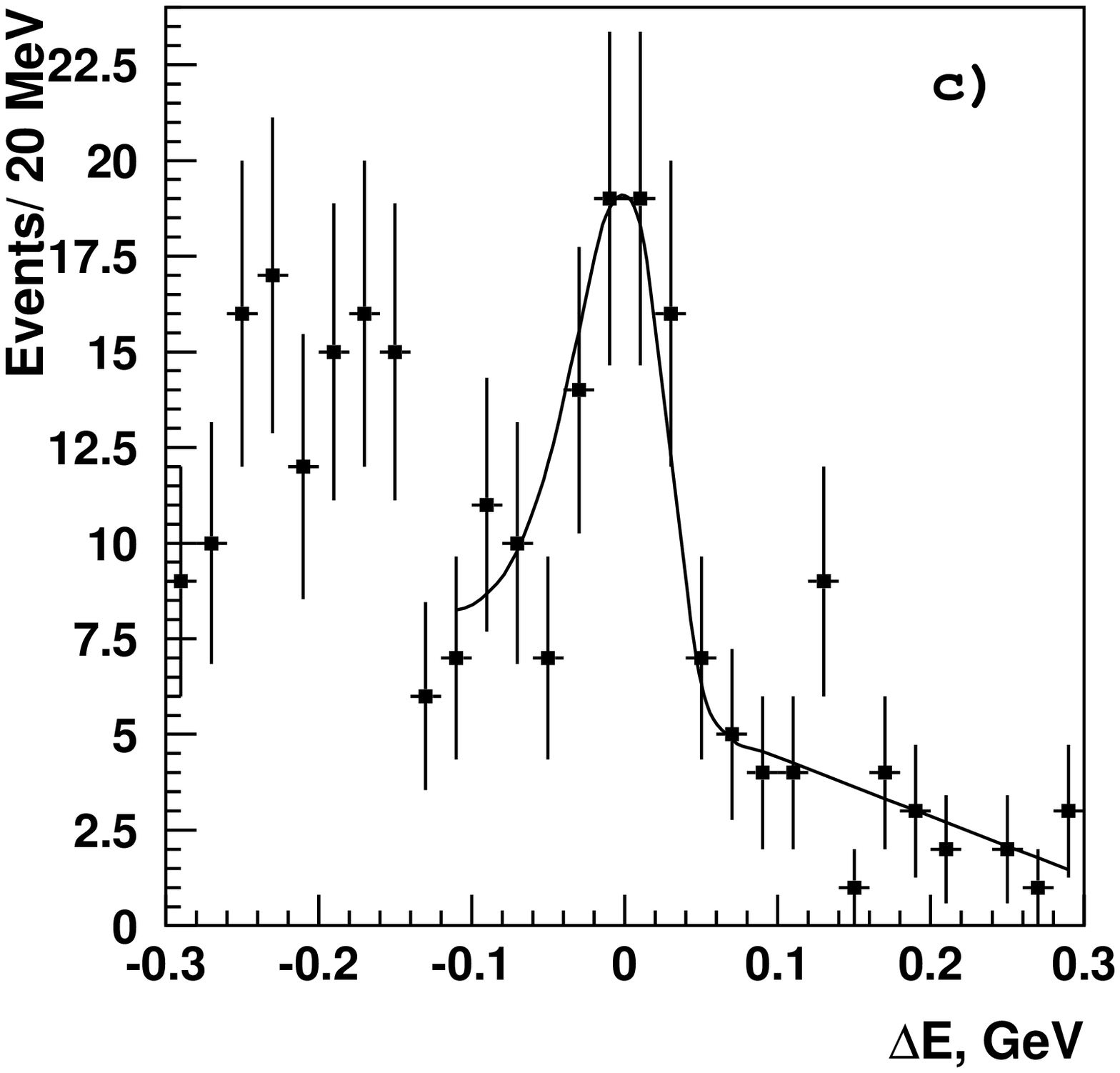}
\includegraphics[width=0.24\textwidth] {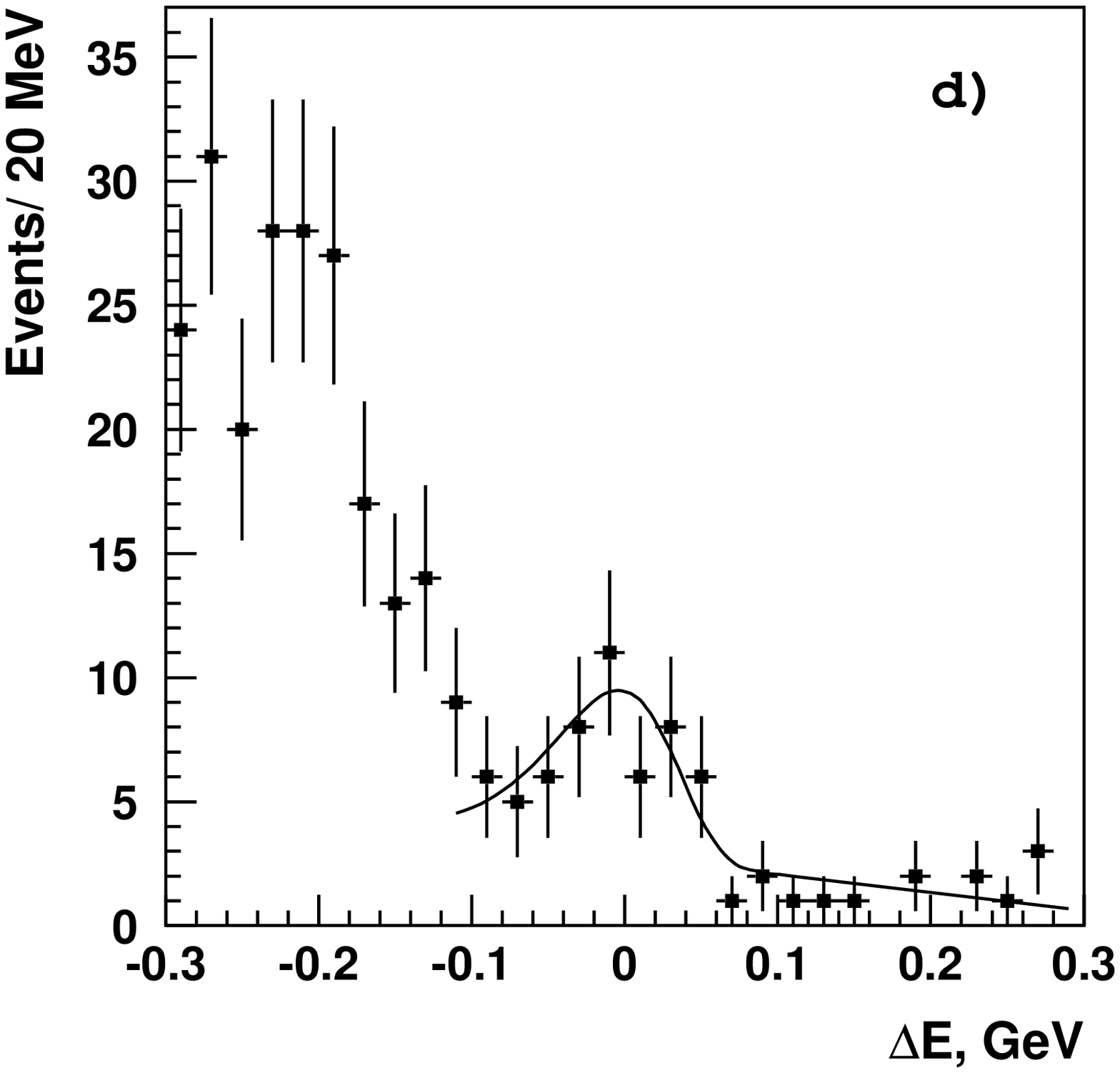}
\caption{$\de$ distributions for the $\bar{B}^0$ decays to 
a) $D\pi^0$, b) $D\omega$, c) $D\eta$ and d) $D^*\pi^0$, $D^*\eta$. 
Points with error bars represent the data 
and curves show the results of the fit.}
\label{de_exp}
\end{figure*}
Signal yields and background levels are determined by fitting 
distributions in $\de$ for candidates in the $\mbc$ signal region.
For each mode, the $\de$ distribution is fitted with an asymmetric
Gaussian for signal and a linear function for background. 
The signal shape is fixed, based on MC simulation.
The region $\de<-0.1$~GeV is excluded from the fit 
to avoid contributions from other $B$ decays.
The results from our fits to the data are shown in Figure~\ref{de_exp} 
and Table~\ref{tabeff}. 
We study the systematic error of the fit by varying the shapes for signal and 
background and changing the fit range. 
The difference in the signal yields does not exceed 5\%. 
We also confirm that there are no feed across between channels 
and other peaking background by using generic $\bb$ MC.
\begin{table}
  \caption{
    \label{tabeff}
    Number of events in the signal region ($N_{\rm tot}$),
    detection efficiency, number of signal events from the
    $\de$ fit ($N_{\rm sig}$)
    and signal purity for the $B\to D^{(*)} h^0$ final states.
  }
  \vspace{0.5\baselineskip}
  \begin{tabular}
    {|l|c|@{\hspace{3mm}}c@{\hspace{3mm}}|@{\hspace{3mm}}c@{\hspace{3mm}}|c|}
    \hline
    Process & $N_{\rm tot}$& Efficiency (\%) & $N_{\rm sig}$  & Purity \\ 
    \hline
    $D\pi^0$ & 265& 8.7 & $157\pm 24$ & 59\% \\
    $D\omega$& 88 & 4.1 &  $67\pm 10$ & 76\% \\
    $D\eta$  & 101& 3.9 &  $58\pm 13$ & 57\% \\
$D^*\pi^0$,$D^*\eta$ & 67 &  &  $43\pm 12$ & 64\% \\
    \hline
    Sum      & 521 &     & $325\pm 31$ & 62\% \\
    \hline
  \end{tabular}
\end{table}

The signal $B$ decay vertex is reconstructed using the $D$
trajectory and the IP constraint. 
The tagging $B$ vertex is obtained with well-reconstructed tracks 
not assigned to the signal $B$ candidate and
the IP constraint~\cite{resol}.
The time difference between signal and tagging $B$ candidates is
calculated using $\Delta t=\Delta z/\gamma\beta c$ and 
$\Delta z = z_{CP}-z_{\rm tag}$.
The proper-time interval resolution function $R_\mathrm{sig}(\dt)$
is formed by convolving four components:
the detector resolutions for $z_{CP}$ and $z_\mathrm{tag}$,
the shift in the $z_\mathrm{tag}$ vertex position
due to secondary tracks originating from charmed particle decays,
and the kinematic approximation that $B$ mesons are
at rest in the CM frame~\cite{resol}.
A small component of broad outliers in the $\Delta z$ distribution,
caused by misreconstruction, is represented by a Gaussian function.
Charged leptons, pions, kaons, and $\Lambda$ baryons
that are not associated with a reconstructed $\bdbkspipi$ decay
are used to identify the $b$-flavor of the accompanying $B$ meson.
The tagging algorithm is described in detail 
elsewhere~\cite{flavor}.

We perform an unbinned time-dependent Dalitz plot fit.
The negative logarithm of the unbinned likelihood function is minimized:
\begin{equation}
 -2\log L= - 2\sum \limits^n_{i=1} \log \{
(1-f_{\rm bg})\, P_{\rm sig}+f_{\rm bg}\, P_{\rm bg}\},
\end{equation}
where $n$ is the number of events.
The function $P_{\rm sig}(m^2_+, m^2_-, \dt)$ is the time-dependent 
Dalitz plot density for the signal events, which is calculated
according to Eq.~(\ref{eq:pdf}) and incorporates reconstruction 
efficiency, flavor-tagging efficiency, wrong tagging probability
and $\dt$ resolution. The function $P_{\rm bg}$ is the probability density 
function (PDF) for the background.
Both $P_{\rm sig}$ and $P_{\rm bg}$ are normalized by
$\int P_{\rm sig, bg}(m^2_+, m^2_-, \dt) dm^2_+ dm^2_- d\dt=1$.
The event-by-event background fraction $f_{\rm bg}(\de, \mbc)$ is 
based on signal and background levels found by fitting $\de$ as described 
above, the $\de$ shape used in the fit, and an $\mbc$ shape that is the 
sum of a Gaussian  signal and an empirical background function with 
kinematic threshold and shape parameters determined from off-resonance data.

\begin{figure}
\includegraphics[width=0.24\textwidth] {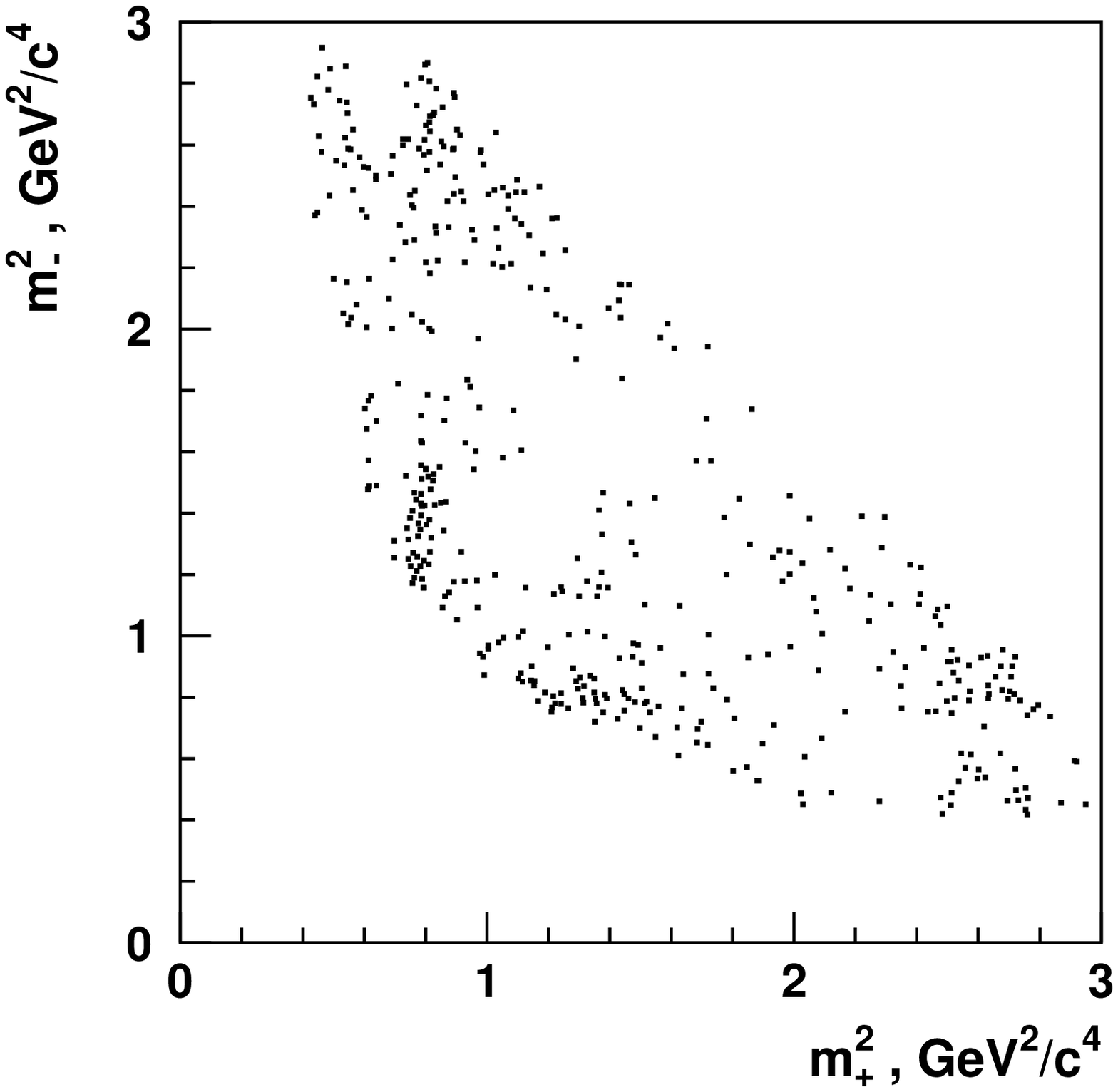}\hfill
\includegraphics[width=0.24\textwidth] {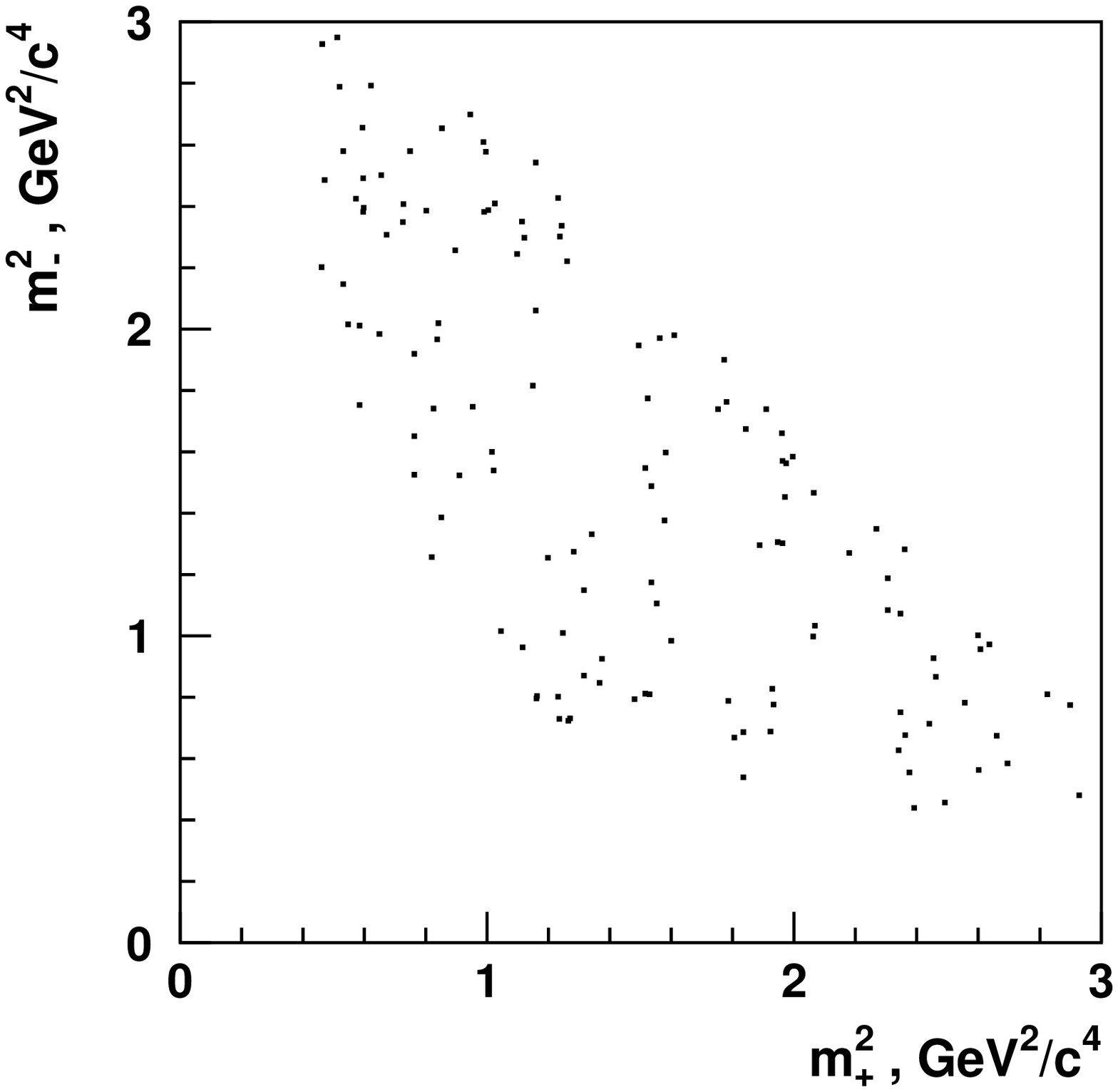}
\caption{Dalitz plot distribution for the $D h^0$ candidates from
  $B$ signal region (left) and $\mbc$ sideband.
}
\label{dalitz}
\end{figure}

We describe the background by the sum of four components: $B$ decays
containing a) real $D$ mesons and b) combinatorial $D$ mesons, and $\qq$
events containing c) real $D$ mesons and d) combinatorial $D$ mesons.
The Dalitz plot is described by the function $f(m_+^2,m_-^2)$ for 
a) and c). 
For b) and d) we use an empirical background
function which includes enhancements near the edges of the Dalitz plot as 
well as an incoherently added $K^*(892)$ contribution~\cite{anton}.
The shape of this function is obtained from an analysis of events in the 
$D$ mass sideband.
The $\dt$ distribution for the $B$ decay backgrounds is described by
an exponential convolved with the detector resolution. For the $\qq$
background, a triple Gaussian form is used, which is obtained from 
events with $|\cos\theta_{\rm thr}|>0.8$.
The use of this sideband region has been validated using MC.
We use the experimental data and generic MC to fix the fractions of
background components.
Figure~\ref{dalitz} shows the Dalitz plot distributions for candidates
in signal and $\mbc$ sideband region, integrated over the entire $\dt$ range 
and $B^0$ and $\bar{B}^0$ combined.
We can see clear differences in these distributions.

The procedure for the $\dt$ fit is tested by extracting  $\tau_{B^+}$ 
using $B^+\to \bar{D}^0[\kspipi]\pi^+$ decay.  We obtain 
$\tau_{B^+}=1.678\pm 0.043$~ps (statistical error only), consistent with 
the PDG~\cite{PDG} value $1.638\pm 0.011$~ps. 

\begin{figure}
\includegraphics[width=0.35\textwidth] {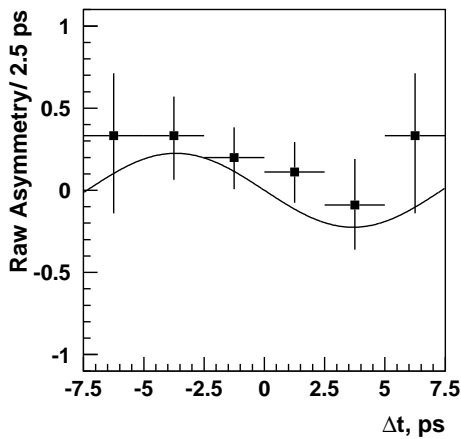}
\caption{Raw asymmetry distribution for the $D^{(*)}[\ks\rho^0]h^0$ 
candidates. The smooth curve is the result of the fit to the full
Dalitz plot.}
\label{asym_data}
\end{figure}

We perform a fit by fixing $\tau_{B^0}$ and $\dm$ at the PDG 
values with a fixed background shape 
and using $\sin 2\phi_1$, $\cos 2\phi_1$ 
as fitting parameters.
The results are given in Table~\ref{fitres} for each of the three final
states separately and for the simultaneous fit over all modes.
\begin{table}
  \caption{
    \label{fitres}
    Fit results for the data. Errors are statistical only.
  }
  \vspace{0.5\baselineskip}
  \begin{tabular}
    {|l|c|c|}
    \hline
Final state & $\sin 2 \phi_1$ & $\cos 2 \phi_1$\\ \hline
$D\pi^0$, $D\eta[\gamma\gamma]$  & 
     $0.80^{+0.54}_{-0.60}$ & $2.07^{+0.78}_{-0.91}$ \\
$D\omega$, $D\eta[3\pi]$ & 
      $0.43\pm 0.90$ & $1.53^{+0.67}_{-0.93}$ \\ 
$D^*\pi^0$,$D^*\eta$ & 
      $1.07\pm 1.14$ & $3.46^{+1.80}_{-2.01}$ \\ 
Simultaneous fit & 
$0.78\pm 0.44$ & $1.87^{+0.40}_{-0.53}$ \\
    \hline
  \end{tabular}
\end{table}
We check goodness-of-fit using one dimensional projections to $\ks\pi^\pm$ 
and $\pi^+\pi^-$ invariant masses and $\delta t$ and find no pathological 
behaviour.
To illustrate, 
the raw $CP$ asymmetry distribution for $D^{(*)}h^0$ candidates with an 
additional constraint $|M_{\pi^+\pi^-}-0.77|<0.15$~GeV$/c^2$, to select 
events consistent with $D\to\ks\rho$, is displayed in Fig.~\ref{asym_data}.
For $D^*h^0$ candidates we take into account the opposite $CP$ asymmetry.
In this case the system behaves approximately as a $CP$ eigenstate, 
with an asymmetry proportional to
$-\sin 2\phi_1$.  

Uncertainty of the $\dnkspipi$ decay model is
one of the main sources of systematic error for our analysis. 
We repeat the fit using two additional decay models from 
CLEO~\cite{dkpp_cleo} and similar Belle analysis~\cite{anton_new}.
The difference between these models and our primary model~\cite{anton} 
is in describing of wide resonances in $\pi^+\pi^-$ and $\ks\pi$, 
non-resonant part and doubly Cabibbo suppressed channels.
The CLEO~\cite{dkpp_cleo} does not include
wide resonances $\sigma(600)$ and $f_0(1370)$, and the doubly Cabibbo 
suppressed channel $D^0\to K^{*+}(1430)\pi^-$.
Another Belle model~\cite{anton_new} has an additional contributions from
$K^*(1410)^\pm pi^\mp$ and $\ks\rho^0(1450)$.
The difference in fitted values for $\sin 2\phi_1$ and $\cos 2\phi_1$
between the nominal model~\cite{anton} and others is found not to exceed
0.1, and we assign this value as a model uncertainty.

We vary the background descriptions to estimate the systematic
uncertainty due to the background parameterization.
We use only a combinatorial and only a signal $D$ PDF for the Dalitz plot 
distribution.
For the time dependence, we consider cases with only a $\qq$ component
or only a $\bb$ component.
The differences 
do not exceed 0.2 and we take this value as a systematic error.

Other contributions to the systematic error are found to be small:
vertexing and flavor tagging (0.02), neglecting suppressed amplitudes
(0.01), signal yield determination (0.02).

The measurement of $\sinres$ is consistent with the high statistics 
measurement in the $J/\psi K^0$ channel~\cite{PDG}. 
The result of $\cosres$ allows one to distinguish between two
solutions in $\phi_1$: $23^\circ$ and $67^\circ$.
We define the confidence level at which the $67^\circ$ solution
(negative value of $\cos 2\phi_1$) can be excluded as
$CL(x)=f_+(x)/(f_+(x) + f_-(x))$, where
$f_+(x)$ ($f_-(x)$) is the likelihood to obtain the fit result 
$\cos 2\phi_1=x$ when true $\cos 2\phi_1$ value of $0.689$ ($-0.689$).
To evaluate $f_+$ and $f_-$ we use sample of 2500 pseudo-experiments 
with the same size as data for both hypotheses. 
We fit these distributions with a sum of two Gaussians.
We calculate $CL$ for $x=1.87$, $1.55$ and $2.09$ to
take into account systematic uncertainties of $^{+0.22}_{-0.32}$
in our $\cos 2\phi_1$ measurement.
As a final result we use the smallest value $CL(1.55)=98.5\pm 0.2\%$, 
excluding the $67^\circ$ solution at $\clres$ C.L.

In summary, we have presented a new method to measure the Unitarity
Triangle angle $\phi_1$ using a time-dependent amplitude analysis of 
the $D\to\kspipi$ decay produced in the processes 
$\bar{B}^0 \to D^{(*)}h^0$. 
We find $\sinres$ and $\cosres$. 
The sign of $\cos 2\phi_1$ is determined to be positive at $\clres$ C.L.,
strongly favoring the $\phi_1=23^\circ$ solution.

We thank the KEKB group for excellent operation of the
accelerator, the KEK cryogenics group for efficient solenoid
operations, and the KEK computer group and
the NII for valuable computing and Super-SINET network
support.  We acknowledge support from MEXT and JSPS (Japan);
ARC and DEST (Australia); NSFC and KIP of CAS 
(contract No.~10575109 and IHEP-U-503, China); DST (India); 
the BK21 program of MOEHRD, and the
CHEP SRC and BR (grant No. R01-2005-000-10089-0) programs of
KOSEF (Korea); KBN (contract No.~2P03B 01324, Poland); MIST
(Russia); ARRS (Slovenia);  SNSF (Switzerland); NSC and MOE
(Taiwan); and DOE (USA).

\end{document}